\documentclass[aps,12pt,float]{revtex4}
\usepackage{graphicx}
\usepackage[english]{babel}
\usepackage{amsmath}
\usepackage{amssymb}

\newcommand{\be}{\begin{eqnarray}}
\newcommand{\ee}{\end{eqnarray}}

\begin{document}

\title{Specific Heat and Sound Velocity Distinguish the Relevant Competing Phase in the Pseudogap Region of High Temperature Superconductors}
\author{Chandra M. Varma$^{1}$ and Lijun Zhu$^{1}$}
\affiliation{$^{1}$Department of Physics and Astronomy, University of California, Riverside, CA 92521, USA }
\maketitle

{\bf A great step forward towards the understanding of high temperature superconductors are the variety of experimental results which have led to the wide-spread acceptance of the idea \cite{cmv1997}  that a phase with a broken symmetry competes with superconductivity in the under-doped region, often called the pseudo-gap region. There are a plethora of suggested phases \cite{Comin2014,Tranquada1996,Kim2008,Chang2012,ghiringhelli,xray2,xray3,Achkar2012,Torchinsky2013,Hoffman2002,Li2006,simon-cmv,Fujita2014a,Fujita2014b}. 
The idea, that a broken symmetry phase competes with superconductivity makes thermodynamic  sense only if the energy gained due to it is comparable to that gained through the superconducting transition in their co-existence region. Extraordinarily, however, no specific heat signature of a phase transition has been identified at the pseudo-gap temperature $T^*$. 
We use the recent highly accurate sound-velocity measurements \cite{Shekhter2013} and the best available specific heat measurements in YBa$_2$Cu$_3$O$_{6+\delta}$ \cite{loram-jpcs,loram-prl1993, Cooper2014} to show that phase transitions to the universality class of the loop-current ordered state with free-energy reduction similar to the measured superconducting condensation are consistent with the sound velocity and with lack of identifiable observation in the specific heat. We also compare the  measured specific heat with some more usual transitions and show that transitions with such symmetry classes can easily be shown by existing specific heat measurements to have energy reduction due to them less than 1/20 the superconducting condensation energy.
}

Sound velocity changes near any phase transition, as shown below, are proportional to $\gamma(T) = C_v/T$, where $C_v$ is the specific heat. Since they are measured with a factor of $O(10^{-2})$ greater accuracy than the deductions from the best available specific heat measurements, they can be utilized to decipher the universality class to which the transition at $T^*$ belongs and therefore their specific heat. 
The free-energy reduction can be calculated from the specific heat expected due to the broken symmetry with the condition that it is similar to the condensation energy measured through the specific heat experiments . 

The free-energy due to a phase transition at a temperature $T_x$ is a homogeneous function of $(T-T_x)$. The elastic constants of a solid are given by the second-derivative of the Free-energy with respect to the relevant strain. Therefore the  isothermal sound velocity variation $\delta c_{\lambda}$ in a polarization $\lambda$ associated with the phase transition, normalized  to the background smoothly varying sound velocity $c_{0 \lambda}$  for  $\delta c_{\lambda} \ll c_{0 \lambda}$ is given by
\be
\label{sound-spht}
\frac{\delta c_{\lambda}(T-T_x)}{c_{0 \lambda}} =\frac{1}{2\rho c_{0 \lambda}}\left[ -\frac{C_v(T-T_x)}{T} \left(\frac{dT_x}{du_{\lambda}}\right)^2 + S(T-T_x)\frac{d^2T_x}{du_{\lambda}^2}\right] .
\ee
$C_v(T-T_x)$ and $S(T-T_x)$ are the specific heat at constant volume and entropy associated with the part of the free-energy associated with the transition at $T_x$, i.e. the part which is a homogeneous function of $(T-T_x)$. $\rho$ is the density. In mean-field phase transitions, such as the superconducting transition, this reduces to the relation commonly used.
Noting that the second contribution above is much smoother than the first and typically $(1/T_x)\big(\frac{dT_x}{du_{\lambda}}\big)^2$ is similar or larger than $\frac{d^2T_x}{du_{\lambda}^2}$, we need to consider only the first term.
Comparing the sound velocity variations at two different transitions, one at $T_c$ and the other at $T^*$, 
\be
\label{relsound}
\frac{\delta c_{\lambda}(T-T^*)}{\delta c_{\lambda}(T-T_c)} = \frac{C_v(T-T^*)}{C_v(T-T_c)} \left(\frac{dT^*/du_{\lambda}}{dT_c/du_{\lambda}}\right)^2.
\ee

\begin{figure}
\centerline{\includegraphics[width=0.8\textwidth]{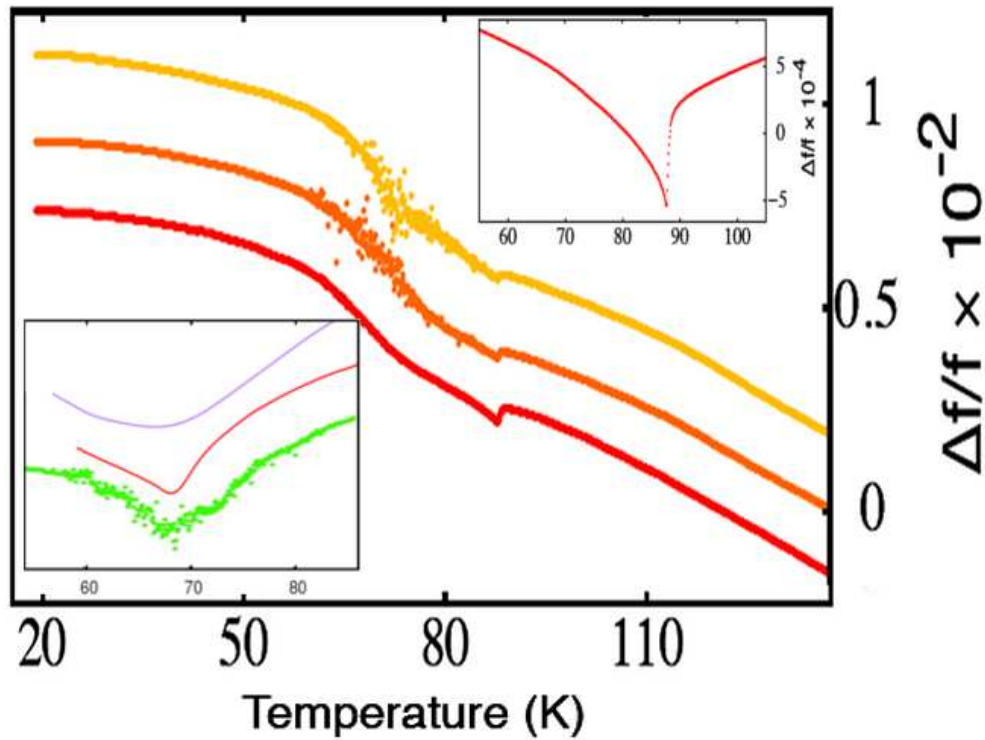}}
\caption{The temperature variation of the relative frequency changes or sound velocity changes measured by Shekhter et al., taken from Fig. 2b of Ref. \onlinecite{Shekhter2013}. Three different modes are shown with the vertical axis displaced. The upper and lower insets show amplified regions near the superconducting and pseudogap transitions with smooth background subtracted, taken from Fig. 1c and Fig. 4c (green curve) of Ref. \onlinecite{Shekhter2013}, respectively. The temperature dependence of the variation in sound velocity for the AT model, using Eq. (\ref{sound-spht}) for two different sets of parameters in Fig. (\ref{gamma-atmodel}), are also shown in the lower inset. The three curves in the inset are mutually displaced. To obtain these results from the dimensionless temperature scale in Fig. (\ref{gamma-atmodel}),  the peaks are made to coincide with the experimental transition temperature of about 68 K and with a multiplicative factor for the vertical scale required by Eq. (\ref{sound-spht}). This multiplicative factor is specified in the text. Note the sharpness of the signature of the superconducting transition in the sound velocity and the wide fluctuation region at the lower transition which is reproduced by the two theoretical curves which bound the parameters needed to fit the experimental results.} 
\label{Shekhter-sound}
\end{figure}

Changes in sound velocity, consistent in their location in the phase diagram with the previous observations both of $T_c$ and $T^*$, have been measured \cite{Shekhter2013} at two dopings in very high quality samples of 
 YBa$_2$Cu$_3$O$_{6+\delta}$.
The results for the larger doping, $\delta = 0.98$, which is close to the peak of the superconducting transition temperature, and close to the putative quantum-critical point, are particularly interesting since the multiplicative factor in Eq. (\ref{sound-spht}) is amplified near the quantum-critical point.  Fig. (\ref{Shekhter-sound}) gives the measured change in relative frequency for three different modes, or equivalently, the relative change in the sound velocity as a function of temperature \cite{Shekhter2013} showing both the  superconducting transition and a transition at about 68 K consistent with the continuation of the loop-current order transition seen through polarized neutron scattering \cite{Bourges-rev}. The insets gives on a different scale the signature of the superconducting transition and the pseudo gap transition after subtracting the background signal. 
The width of the transition at the d-wave superconducting transition at $T_c$ is measured to be about 0.1 K~\cite{Ramshaw-pc}, testifies to the quality of the samples used. It also means that the  signature at the lower transition may not be attributed to any disorder which leads to broadening a d-wave  superconducting transition more than about 0.1 K.

\begin{figure}
\centerline{\includegraphics[width=0.9\textwidth]{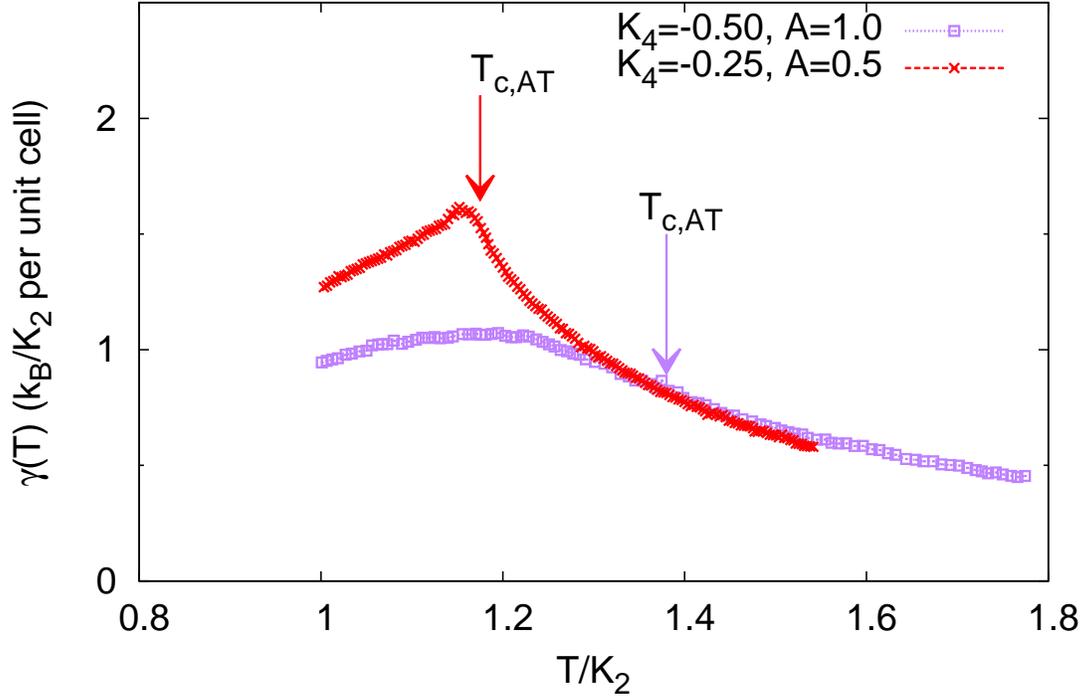}}
\caption{The exactly calculated specific heat coefficient $\gamma(T)$ for the Ashkin-Teller model  for two different values of the parameters from Gronsleth et al. \cite{Gronsleth}. The temperature scale is normalized to the parameter $K_2$ of the AT model. The transition temperature in terms of this scale is given in that Reference and is close to but does not coincide with the peak of the specific heat, but marks the singularity of the order parameter. The vertical scale in the figure is such that the asymptotic high temperature entropy, $S(T) = \int_0^T dT' \gamma(T')$, for all the curves is $k_B \ln 4$/unit-cell, corresponding to a model of 4 unit-vectors per unit-cell. 
}
\label{gamma-atmodel}
\end{figure}

\begin{figure}
\centerline{\includegraphics[width=0.9\textwidth]{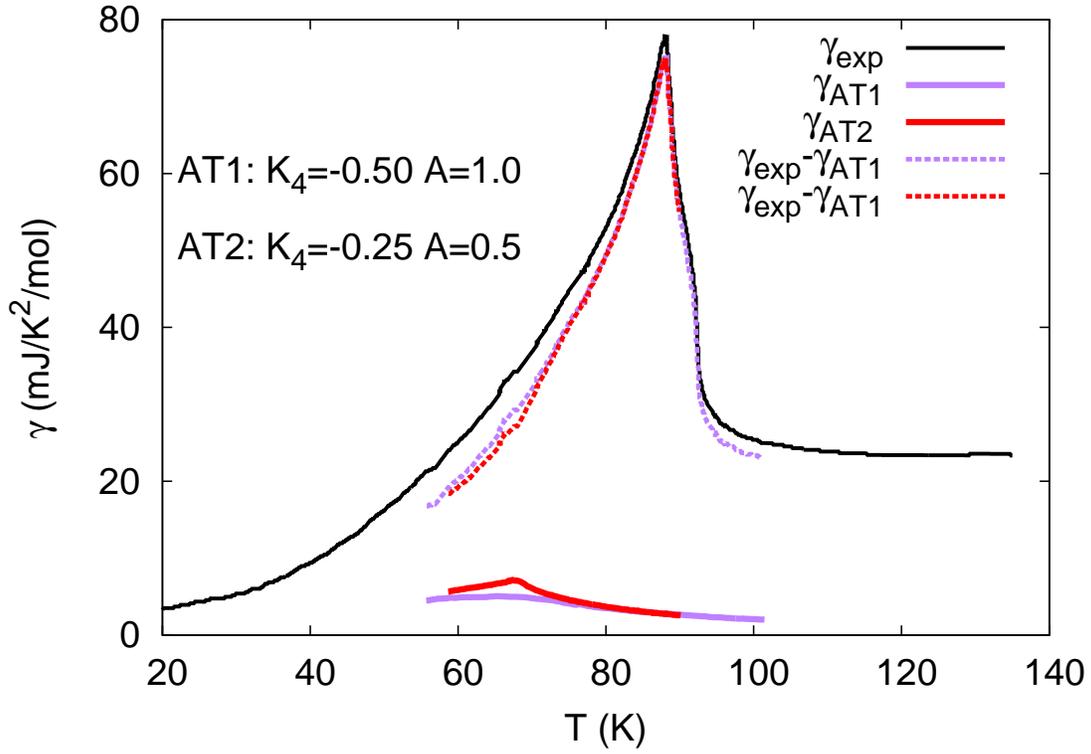}}
\caption{Comparison between the measured $\gamma(T)$ from Cooper et al. \cite{Cooper2014} and $\gamma(T)$ calculated for the AT model with two sets of parameters. The latters have the same condensation energy as the superconducting condensation energy. Also shown are their differences $\gamma_{exp}-\gamma_{AT}$, providing an estimation of $\gamma(T)$ if there were superconducting transtion only. The physics of why the small $\gamma(T)$ of the AT model gives the same condensation energy as the mean-field superconducting condensation energy is easily understood from the wide and relatively smooth fluctuation regime of its entropy and the fact that the specific heat is related to the derivatives of the entropy [see Fig. (\ref{Entropy}) in Supplementary Information].} 
\label{RelGamma}
\end{figure}

Let us first consider whether the thermodynamical properties of the loop-current order are consistent with the sound-velocity (and specific heat) measurements. This broken symmetry, with which many varieties of experiments \cite{Bourges-rev, Greven, Kapitulnik1, Kapitulnik2,Leridon, kaminski, Armitage, Shekhter2013} are consistent, is in the statistical mechanical class 
of the Ashkin-Teller (AT) model which does not have a specific heat divergence but singularities exist in the order parameter as a function of temperature. The asymptotically exact critical exponents were derived by Baxter \cite{Baxter} - over the range of the parameters for the AT model consistent with the symmetry of the observed phase, the specific heat exponent varies from 0 to - 2. The specific heat as a function of temperature (and the order parameter) was calculated \cite{Gronsleth} by Monte-Carlo methods on asymptotically large lattices and are given for two sets of parameters in Fig.(\ref{gamma-atmodel}).  In the lower inset of Fig. (\ref{Shekhter-sound}), the sound-velocity change near $T^* \approx 68 K$ is plotted together with the prediction from the AT model for this two sets of parameters.  These parameters were chosen among those calculated \cite{Gronsleth} to be such as to give, using Eq. (\ref{sound-spht}), a sound velocity signature similar to those observed. Note that the two theoretical curves shown bound the region of parameters which reproduce the sound velocity variations, as well as 
the striking similarity between the measurements and the calculations with regard to the wide fluctuation regime for transitions of this universality class. 
Having thus approximately fixed the parameters of the AT model, we calculate the specific heat quantitatively to compare with the observations. The free-energy reduction due to the transition(s) is calculated from the measured specific heat \cite{loram-jpcs,loram-prl1993,Cooper2014} [see Supplementary Information for details]. The value obtained is 52.7 Joules/mole. The calculated specific heat  for the AT model \cite{Gronsleth} shown in Fig. (\ref{gamma-atmodel}) is for the four discrete states of the AT model with one classical spin 1 per unit-cell, so that the asymptotic entropy at high temperatures in every case is $k_B\ln 4/$unit-cell. We must determine the value of the effective spin for the ordered state. Like in any transition in which the relevant free-energy is drawn to collective degrees of freedom from that of itinerant Fermions, this can be an arbitrary number. But if the transition is as significant thermodynamically as the superconducting transition, we can determine it by calculating the reduction in  energy due to it to approximately equal the superconducting condensation energy.  The procedure is also described in Supplementary Information. With this constraint $\gamma(T)$ for the two sets of parameters of the AT model is plotted together with the experimental $\gamma(T)$ in Fig. (\ref{RelGamma}). We present also the measured value minus the calculated $\gamma(T)$. It is obvious due to the wide fluctuation region that it is not possible to tell from the specific heat that there is the phase transition near 68 K visible in the sound velocity measurement. This is even without taking into account the error bars in the specific heat measurements.

The peak of $\gamma(T)$ in the two theoretical curves are 6.1 and 4.1 mJoules/(mole K$^2$), a factor of 6 and 9 below the measured value, respectively.  The width of the peak at half-height is about 40 K. We can now try to estimate the error bars in the specific heat measurements. This has not been given in the papers reporting the deduction of the electronic specific heats \cite{loram-jpcs,loram-prl1993,Cooper2014}. We provide the minimal estimates of errors in Supplementary Information and find it to be between 1.5 mJ/(mole K$^2$). 
Note that the experimental specific heat feature as shown in Fig. (\ref{RelGamma}) gives evidence for multiple superconducting transitions in a width of about 5 K. Given the width of the specific heat expected in the pure limit and the error bars of measurements and the width of the transition, even an uncertainty in determining $\gamma(T)$ of 0.5 mJoule/(mole K$^2$) would not allow the specific heat bump for the smoother of the two sets of parameters to be decipherable.

It should be mentioned that a broad bump in the specific heat of the magnitude expected from the AT model is claimed to be observed \cite{momono} for La$_{2-x}$Sr$_x$CuO$_4$, systematically decreasing to lower temperature as doping is increased and invisible above $x = 0.22$.   Ultrasound anomalies, besides those at the superconducting transition temperature, in this compound, as well as those in YBa$_2$ Cu$_3$O$_{6+\delta}$ with a $T_c \approx 85 K$ at what we would now call $T^* \approx 120 K$, albeit not so clear-cut as the latest measurements, were seen long ago \cite{Shobho}.

Let us return now to Eq. (\ref{relsound}) to estimate the relative factors of variation of the transition temperatures with strain.
From Fig. (\ref{RelGamma}), the peak height of $\gamma(T)$ at near $T^*$ may be taken to be about a factor of 8 smaller than the peak height of $\gamma(T)$ near $T_c$. The maximum value of  the ultrasound anomaly ($\Delta f/f$) of Fig. (\ref{Shekhter-sound}) has a typical value of $7\times 10^{-4}$ at $T_c$, while that around $T^*$ has a maximum value above background of $1.2\times10^{-3}$ \cite{Ramshaw-pc}. Using Eq.(\ref{relsound}), one concludes that  $dT^*/du_{\lambda}$ should be be about 4 times larger than $d T_c/du_{\lambda}$. The modes are mixtures of different polarizations for each of which we do not have separate information. We only know that the volume of the crystal changes linearly with $\delta$ and that near the quantum-critical point $dT^*/d\delta \gg dT_c/d\delta$. So the estimate made here are quite reasonable at least for the build modulus components of the sound velocity.

Let us now contrast the above results to some other classes of phase transitions with respect to the measured specific heat.  Detailed sound velocity measurements, which have been made so far only for a small number of samples, would be even more effective in judging  their relevance as competitors to superconductivity. Some of the transitions are being reported in under-doped cuprates through high quality measurements in NMR \cite{nmr}, neutron scattering \cite{tranquada}, elastic and inelastic x-ray scattering \cite{ghiringhelli,xray2,xray3} and ultrasound measurements in a magnetic field \cite{Proust}. 

Suppose there is a transition in the same class as the superconducting transition, i.e with a nearly mean-field signature.
An examination of the measured $\gamma(T)$ with a "jump" of more than 50 mJoules/mole-K$^2$ of Fig. (\ref{RelGamma}) and an uncertainty in measurement of 1.5 mJoules/mole-K$^2$shows that another mean-field transition would be revealed if its jump in $\gamma$ at $T_x$, is $\gtrsim$ 1/30 $T_c/T_x$.  It would then have at this limit an energy reduction $\lesssim$ 1/30, the superconducting condensation energy. 

Let us consider the Ising model in 2 dimensions. The charge density wave or "nematic" transition, either intra-cell or inter-cell in strictly two dimensions belong to the universality class of the Ising model. We fix the amplitude of "spin" such that the energy reduction is the same as half the measured energy, and calculate $\gamma(T)$ for the exact results known for the Ising model. The resulting $\gamma(T)$ is also shown in Fig. (\ref{RelGamma}) for a $T_x$ chosen to be about 68 K.  One concludes that a transition with a peak $\gamma$  of about 1/20, for the estimate of error bars, i.e. a gain in energy 1/20 of the superconducting condensation energy would have been detectable.  

Even in highly anisotropic electronic materials, the charge density wave transitions are usually three dimensional as evidence by the specific heat signature. A good example is the well studied case of the electronically highly anisotropic material 2H-TaSe$_2$, in which the energy reduction due to the second order charge density wave transition at about 120 K is measured to be about 67 Joules/mole \cite{craven}, only about 15\% larger than the energy reduction deduced from the measured $\gamma(T)$ in YBa$_2$Cu$_3$O$_{6+\delta}$. The exponent in the specific heat has been carefully measured \cite{craven} and is consistent with the 3D-Gaussian value for the Ising model of 0.5.  We plot the measured $\gamma(T)$ due to the transition taken from Fig. (3) of Ref.[\onlinecite{craven}] scaled to half the measured condensation energy in YBa$_2$Cu$_3$O$_{6+\delta}$ also in Fig. (\ref{RelGamma}) at the transition temperature of 2H-TaSe$_2$. It is clear that a structural transition of this type with about (1/45) of the condensation energy of the superconducting transition, would be easily visible in the specific heat measurements of Loram et al. if it were to occur at 120 K and proportionately smaller fraction of condensation energy would be detectable at lower temperatures. Appeal to disorder for the 2-D Ising case or this case is not of much help; the broadening of the transitions due to disorder would have to be more than about 20 K, much larger than evident in the measurements, for them to have escaped attention. Sound velocity measurements on high quality samples can put much stronger bounds on their relevance. For dopings such that  $T^* > T_c$, they can also be similarly excluded because their specific heat signatures would have to be even larger.

If the energy gain due to other transitions is much less than the superconducting condensation energy, they may be regarded as phenomena incidental to the principal remarkable features of the phase diagram of cuprates which are all extravagant in spending the free-energy. As mentioned a variety of charge stripe/checkerboard phases \cite{Comin2014,Tranquada1996,Kim2008,Chang2012,ghiringhelli,xray2,xray3,Achkar2012,Torchinsky2013,Hoffman2002,Li2006,Fujita2014a,Fujita2014b} with varying support in experiments have been discussed without their magnitudes from which the energy reduction may be estimated. They are all of the Ising/CDW variety discussed in relation to Fig. (\ref{RelGamma}). An exceptional case is the intra-unit-cell ${\bf Q} =0$ charge order \cite{Fujita2014a,Fujita2014b} of
d$_{x^2-y^2}$ symmetry deduced by two very refined experiments. If this order parameter were to occur independently, it also belongs to the Ising class and can be ruled as irrelevant. But such a distortion has been shown \cite{Shekhter2009} by symmetry considerations to be mandated in the loop-current ordered state and to be proportional to the modulus square of the loop order parameter. We suggest further experiments to study the temperature dependence of the distortion to check whether this occurs together with loop-current order. If so, and if they are of small magnitude, the charge density observations of this class \cite{Fujita2014b}, the direct observations of the loop-current order \cite{Bourges-rev} and its manifestations in a variety of other experiments \cite{Greven, Kapitulnik1, Kapitulnik2,Leridon, kaminski, Armitage}, the specific heat \cite{loram-jpcs,loram-prl1993} and the sound velocity \cite{Shekhter2013} are all mutually consistent.

{\bf Acknowledgements}:  \\
The authors wish to thank B. Ramshaw and A. Shekhter for discussions of the sound velocity measurements. This research was supported by National Science Foundation grant DMR 1206298.


\newpage

\clearpage

\clearpage

\clearpage

\begin{figure}
\centerline{\includegraphics[width=0.9\textwidth]{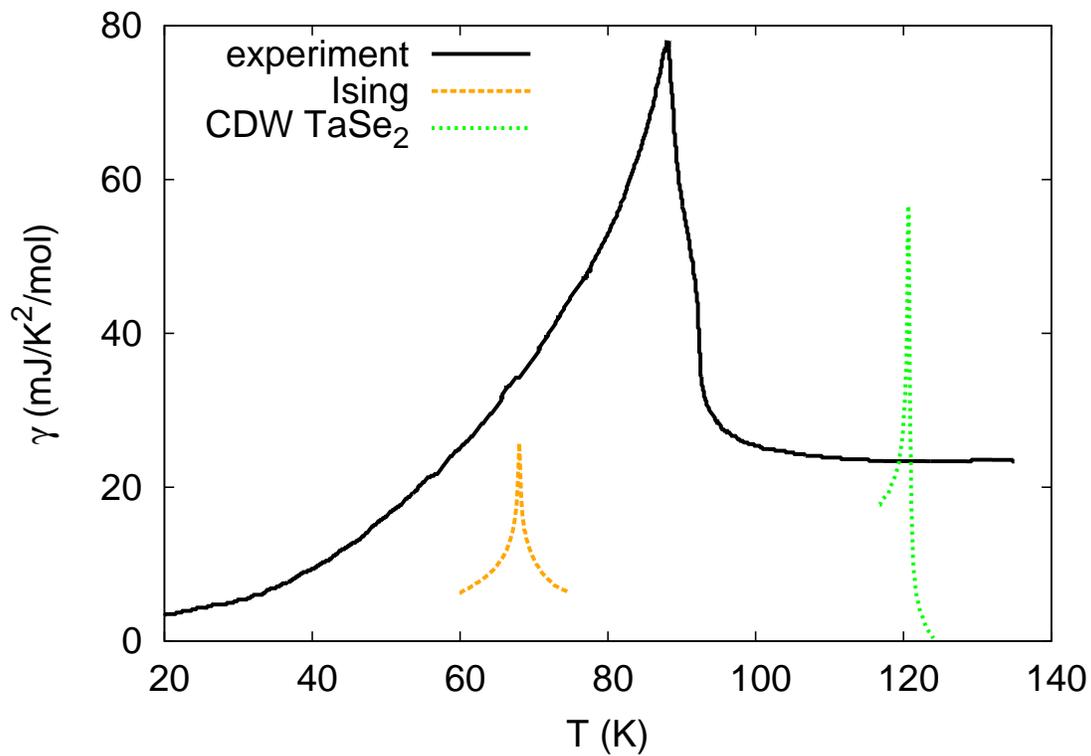}}
\caption{Comparison between the measured $\gamma(T)$ from Cooper et al. \cite{Cooper2014} and $\gamma(T)$ calculated for the 2D Ising model,  and 2H-TaSe$_2$ with CDW transition\cite{craven}.} 
\label{Compare}
\end{figure}
\clearpage


\begin{center}
{\bf {\Large Supplementary Information}}
\end{center}

\section{Calculations of the condensation energy}

A thermodynamically exact procedure to calculate the reduction in energy due to any phase transition at a temperature $T_x$ is
\begin{equation}
\label{E}
\Delta E = \int_0^{T_{\infty}}dT [S_{n}(T) - S(T-T_x)].
\end{equation}
Here $S(T-T_x)$ is the measured  entropy due to the transition at $T_x$ and $S_{n}(T)$ is the entropy for the relevant disordered state, i.e. the entropy for the state at all temperatures with the coupling constant or constants which lead to the particular transition set to zero. The upper integral limit $T_{\infty}$ is chosen as the temperature where the two entropies become immeasurably the same. Eq.(\ref{E})  also takes into account the fluctuation contribution to the reduction in energy. For superconducting transition, $S_{n}(T)$ is that of the Fermi-liquid; for mean-field phase transitions like superconductivity, $T_{\infty}$ in Eq. (\ref{E} ) is close to $T_c$ because there is hardly any fluctuation contribution. This gives the customary condensation energy given for the superconducting transition.

 \begin{figure}
\centerline{\includegraphics[width=0.9\textwidth]{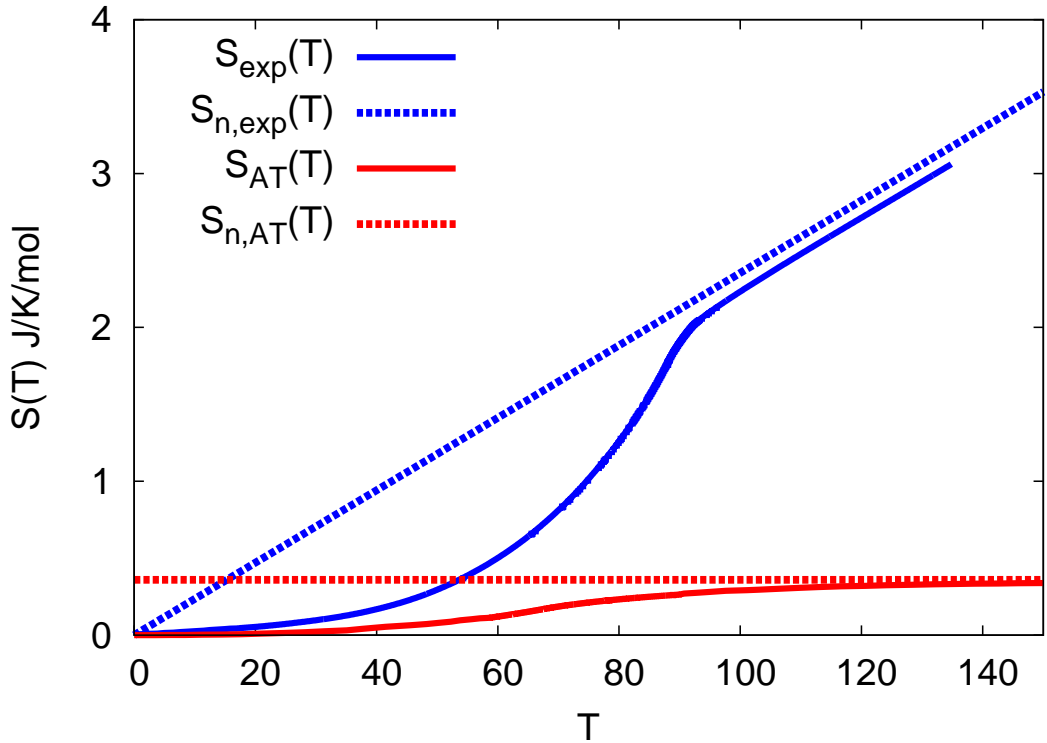}}
\caption{The experimental entropy- solid blue line and the entropy from the AT model, solid red line as a function of temperature,
with the scale adjusted for the latter so that the condensation energy from it alone is equal to 1/2 the measured condensation energy below $T_c$. The condensation energy for the AT model is the areas between the solid and the dashed red curves. The dashed curve is the entropy for the un-ordered phase of the AT model. The energy gain in the experiment is calculated from the difference of the experimental curve and a straight line extrapolating the linear part of this curve above $T_c$  to $T=0$. The dashed blue curve is the extrapolation of the 
deduction of entropy of the normal metallic state from the measured constant $\gamma$ from above $T_c$. The difference of the two blue curves is used to estimate error bar in the specific heat measurements as discussed in the text. } 
\label{Entropy}
\end{figure}

Fig. (\ref{Entropy}) converts the measured $\gamma(T)$ \cite{Cooper2014} to the entropy $S(T)$ as a function of temperature. Then a calculation using Eq. (\ref{E}), the energy gained due to the transition(s) in YBa$_2$Cu$_3$O$_{6.98}$ is 52.7 Joules/mole.

Next, we calculate the reduction in energy of the AT model using the specific heat shown in Fig (2) of main text by the same procedure with a transition temperature $T^* \approx$ 68K and with an overall multiplicative factor $p$ which must be determined. $p$ is the effective value of the square of the ordered moment for the representation of the physics of the transition by the AT model. From neutron scattering experiments, it is roughly estimated to be $O(0.1)^2$ per unit-cell, counting the two staggered orbital moments in a two-dimensional cell of about 0.05 $\mu_B$. This was used earlier \cite{Gronsleth} to discuss the observability of the transition through specific heat measurements. But there is a much better way of estimating $p$. We calculate the condensation energy from the $\gamma(T)$ calculated from the AT model using Eq.(\ref{E}) and equate it to 1/2 the experimentally measured reduction in energy which as we have noted above is  $\approx 52.7 Joules/mole$. The factor of about (1/ 2) is required in the argument, because the measured condensation energy includes the contributions both of superconductivity and of the pseudo-gap formation. 

\begin{figure}
\centerline{\includegraphics[width=0.9\textwidth]{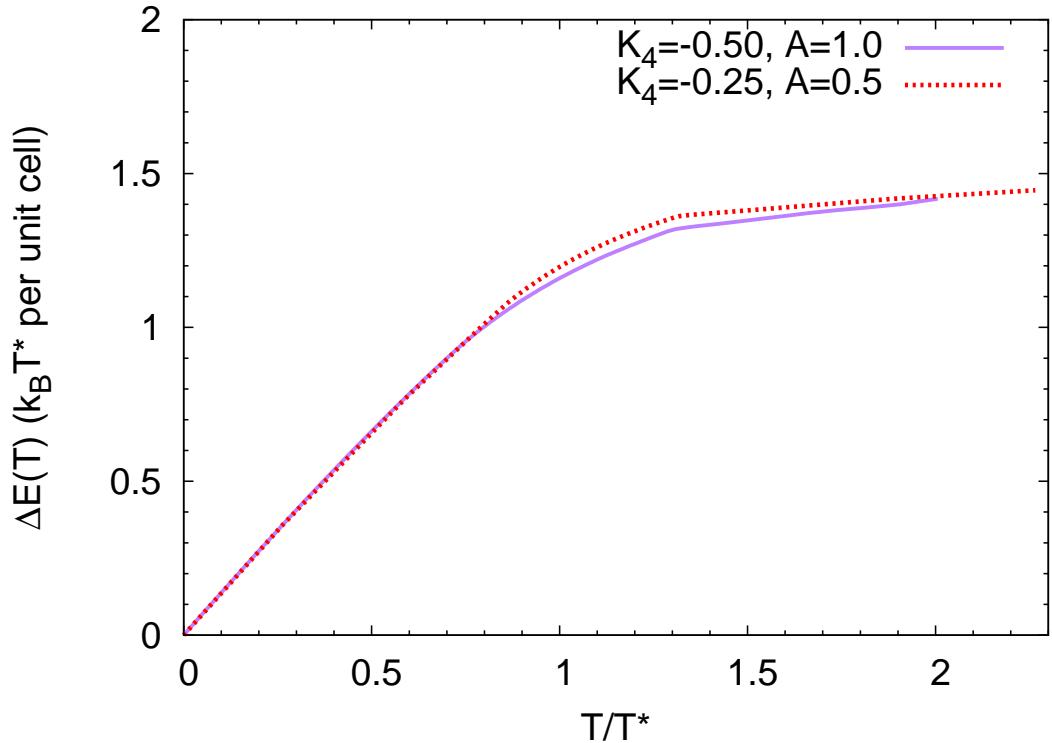}}
\caption{The reduction in  energy of the Ashkin-Teller model calculated from its $\gamma(T)$ for 4 degrees of freedom per unit-cell with unit-moment. $\Delta E(T) = p \int_0^T dT' [S_n(T)-S(T')]$. Here, $p=1$ and  $S_{n}(T) = k_B \ln 4$/unit-cell.} 
\label{DeltaE}
\end{figure}

Fig. (\ref{DeltaE}) gives the total reduction in energy to be $\approx$ 1694 Joules/mole taking into account that a unit-cell has two CuO$_2$ units each with the four discrete degrees of freedom.  This means that the multiplicative factor $p \approx  (1/2)(52.7/1694) \approx$ 1/64 for the reduction in energy of the pseudo-gap to be about the same as  the superconducting condensation energy. $S(T-T_c)$ and $S_{n}(T)$ for the AT model with the same condensation energy are both also shown in Fig. (\ref{Entropy}) together with their superconducting counterparts. Note the smooth shape of the entropy of the AT model compared to experimental curve because most of the former is due to fluctuations peculiar to the AT model. $\gamma(T)$ is related to the derivatives of the curves and for the same condensation energy it would be expected to be much smoother in the former than for superconductivity [see Fig. (3) in main text]. 

\section{Estimate of error in the deduced electronic specific heat}

We can now try to estimate the error bars in the specific heat measurements. This has not been given in the papers reporting the deduction of the electronic specific heats \cite{loram-jpcs}. Recall that the electronic heat capacity, which in the range of interest is about 10$^{-2}$ of the total (lattice plus electronic) specific heat measured, is deduced by subtraction of similar compounds which are insulating. Cooper et al. are to be commended highly that useful information about the condensation energy of the superconducting and the pseudo-gap transition is obtained, supporting the theory and the phase-diagram \cite{cmv1997} that the pseudo-gap boundary $T^*(x)$ marks a distinct phase.  The minimal estimates of the uncertainty may be made as follows. If we continue the constant normal state $\gamma$ to $T\to 0$, the area of the measured curve in Fig.(2) above such a line is about 10\% higher than that below, while thermodynamically they are required to be the same.  This is equivalent to what is shown in Fig. (\ref{Entropy}), where we show in the continuous curve the entropy from the experimental measurement and the entropy from the continuation of the experimental normal state $\gamma$ extrapolated to lower temperatures. The two deviate neat $T_c$ by 0.16 Joules/(mole K).  If we distribute this error equally over the temperature range of measurement, the un-certainty in determination of $\gamma(T)$ is about 1.6 mJ/(mole K$^2$). 
Note that these are the lower limits of the possible error. 

\bibliographystyle{naturemag2}


\end{document}